\documentclass[runningheads]{llncs}
\usepackage{graphicx}
\usepackage{seqsplit}

\begin{document}
\title{Software-Defined Cryptography: \\A Design Feature of Cryptographic Agility}
\titlerunning{Software-Defined Cryptography and Crypto-Agility}
\author{Jihoon Cho \and
Changhoon Lee \and
Eunkyung Kim \and
Jieun Lee \and
Beumjin Cho
}
\authorrunning{J. Cho et al.}

\institute{Security Research Team, Samsung SDS, Republic of Korea \\
\email{\{jihoon1.cho, changhoon47.lee, 
ek41.kim, jieun78.lee, beumjin.cho\}@samsung.com}}
\maketitle

\begin{abstract}

Given the widespread use of cryptography in Enterprise IT, migration to post-quantum cryptography (PQC) is not drop-in replacement at all. Cryptographic agility, or crypto-agility, is a design feature that enables seamless updates to new cryptographic algorithms and standards without the need to modify or replace the surrounding infrastructure.  This paper introduces a notion of software-defined cryptography as the desired design feature for crypto-agility, emphasizing the role of software in providing centralized governance for cryptography and automated enforcement of cryptographic policies, such as migration to PQC.

\end{abstract}

\thispagestyle{empty}

\section{Introduction}

Quantum computers are expected to break conventional public key cryptography once they reach a certain level of performance.  There thus have been efforts to standardize \emph{post-quantum cryptography} (PQC), which offers resistance against attacks enabled by quantum computers.\footnote{https://csrc.nist.gov/projects/post-quantum-cryptography/post-quantum-cryptography-standardization} Considering the widespread use of cryptography in Enterprise IT, however, transition to PQC from conventional public key cryptography is not a drop-in replacement at all. In fact, we have never experienced a full-scale replacement of public key cryptography since its invention due to the pioneering work of Diffie and Hellman in their paper \cite{dif77}, \emph{New Directions in Cryptography}, in 1976.

Rose \emph{et al.\@} \cite{PQC_migration_nature} explored the complexity and strategic requisites involved in this transition, claiming that many information systems cannot adopt new cryptographic algorithms or standards without extensive and time-consuming modification to their infrastructure. Ott \emph{et al.\@} \cite{Ott22} pointed out the lack of related research in literature and questioned whether the applied cryptography and systems research communities have adequately understood and provided a framework for efficient and secure cryptographic transition. Recognizing the complexity of migrating to PQC, the White House issued the National Security Memorandum (NSM-10)\footnote{National Security Memorandum on Promoting United States Leadership in Quantum Computing While Mitigating Risks to Vulnerable Cryptographic Systems.}, directing the National Institute of Standards and Technology (NIST) to launch the ‘Migration to PQC’ project\footnote{https://www.nccoe.nist.gov/crypto-agility-considerations-migrating-post-quantum-cryptographic-algorithms}, inviting industry experts to develop best practices and tools for the migration to PQC. The NSM-10 highlights the importance of \emph{cryptographic agility} in this migration effort, aiming to reduce transition time and facilitate seamless updates for future cryptographic standards. As per the U.S. Department of Homeland Security, cryptographic agility, or crypto-agility, is a \emph{design feature} that allows for agile updates to new cryptographic algorithms and standards without modifying or replacing the surrounding infrastructure.\footnote{https://www.dhs.gov/publication/cryptographic-agility-infographic}

This paper introduces a notion of \emph{software-defined cryptography} as the desired design feature for crypto-agility, underscoring the role of software in order to provide centralized governance for cryptography and realize automated enforcement of cryptographic policies, such as migration to PQC.\footnote{This paper is based on a contributed talk at Real World Crypto (RWC) 2024, titled `Entering a New Era of Crypto Engineering: Cryptographic Visibility and Agility'.} We first investigate software-defined approaches, which have addressed the similar issues by emphasizing the role of software and enabling programmability to centrally manage the behavior of Enterprise IT resources collectively within the operational environment. Recognizing that the transition to PQC is essentially a secure software update, we also examine its best practices, namely DevSecOps, and discuss that cryptographic policies, such as migration to PQC, can be written as a code and automatically enforced mesh in CI/CD pipelines at scale. Finally, we propose a notion of software-defined cryptography, and discuss how service mesh aligns with it.

\section{Design Feature of Software-Defined Approach}

This section investigates how software-defined approach have addressed the issue of attaining governance and automating policy enforcement to manage resources in Enterprise IT. Network solved this problem for the first time, and other domains, such as cybersecurity, has followed by adopting methodologies. We first analyze the design feature of software-defined networks (SDN) and zero trust architecture, and finally discuss how software-defined approach is related to cryptographic agility.

\subsection{Software-Defined Network (SDN)}

\begin{figure}[ht] 
\centering
\includegraphics[width=0.6\linewidth]{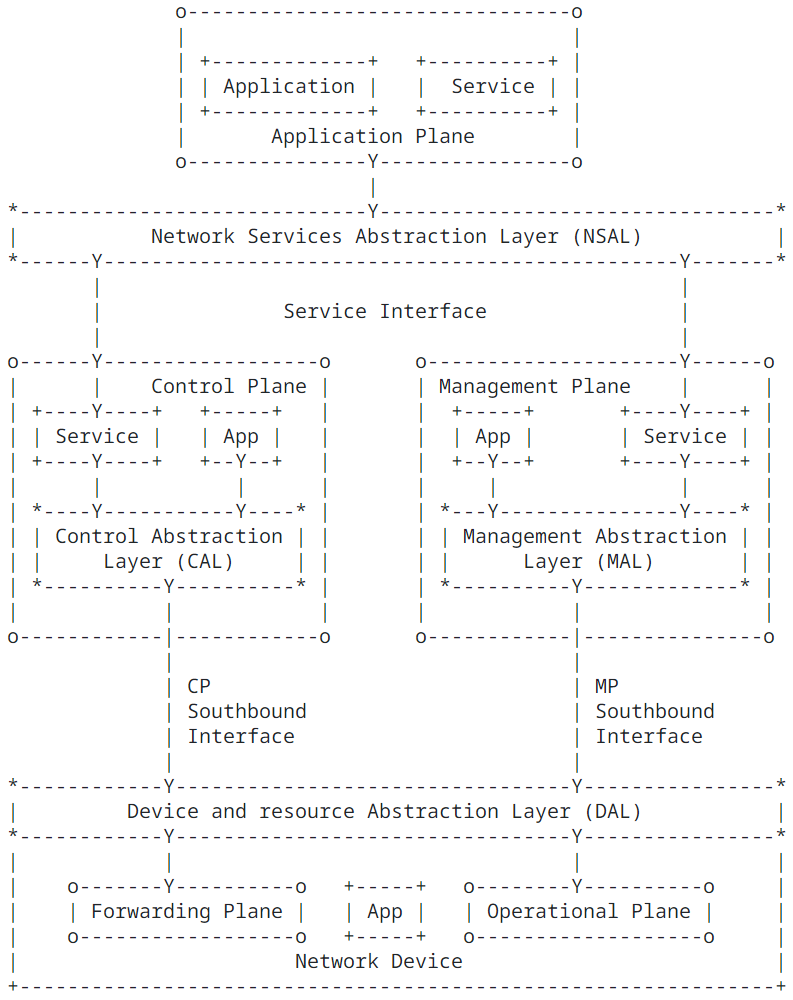}
\caption{SDN Layer Architecture \cite{rfc7426}}
\label{fig:rfc_sdn}
\end{figure}

The advent of Software-Defined Networking (SDN) was primarily driven by the increasing complexity and inflexibility of traditional network architectures. Traditional networks rely heavily on hardware-based devices like routers and switches, which require manual configuration and management. This manual process is not only time-consuming but also prone to errors, making it challenging to scale and adapt to the dynamic requirements of modern IT environments. SDN addresses this issue by introducing \emph{abstractions} between forwarding and control planes, as well as between operational and management planes, as in Figure~\ref{fig:rfc_sdn} \cite{rfc7426}. For example, the \emph{control plane} determines how packets should be forwarded by one or more network devices and implements these decisions on the network devices. The \emph{forwarding plane} handles packets on the data path based on instructions from the control plane.

The abstractions, such as interfaces or APIs, simplify network operations by leveraging software applications to dynamically program policies for individual network devices, thereby enabling \emph{automation} for controlling overall network behavior. The design feature of software-defined networks, which provides interfaces via \emph{abstraction} and enables \emph{automation} via software programming, has been extended to other domains such as software-defined storage, software-defined perimeter, and software-defined data center, resulting in the emergence of the concept known as software-defined everything (SDx).

\subsection{Software-Defined Perimeter and Zero Trust Architecture}

Zero Trust (ZT) is a cybersecurity paradigm that focuses on users, assets, and resources rather than a static, network-based perimeter. Zero Trust Architecture (ZTA) employs these principles to plan enterprise infrastructure and workflows, ensuring authentication and authorization occur before establishing a session with the resource, rather than trusting an asset or user based on physical or network location. However, this is an incredibly challenging task considering complex IT resources in Enterprise IT. Zero Trust Architecture (NIST SP 800-27) \cite{RBMC20}, which outlines the design principles of zero trust, gives insights on how to address such challenges. According to it, ZTA implementations can be achieved using an overlay network, commonly referred to as software-defined perimeter (SDP) approaches, which incorporate concepts, i.e., \emph{abstraction} and \emph{automation}, from SDN.

\begin{figure}[ht] 
\centering
\includegraphics[width=\linewidth]{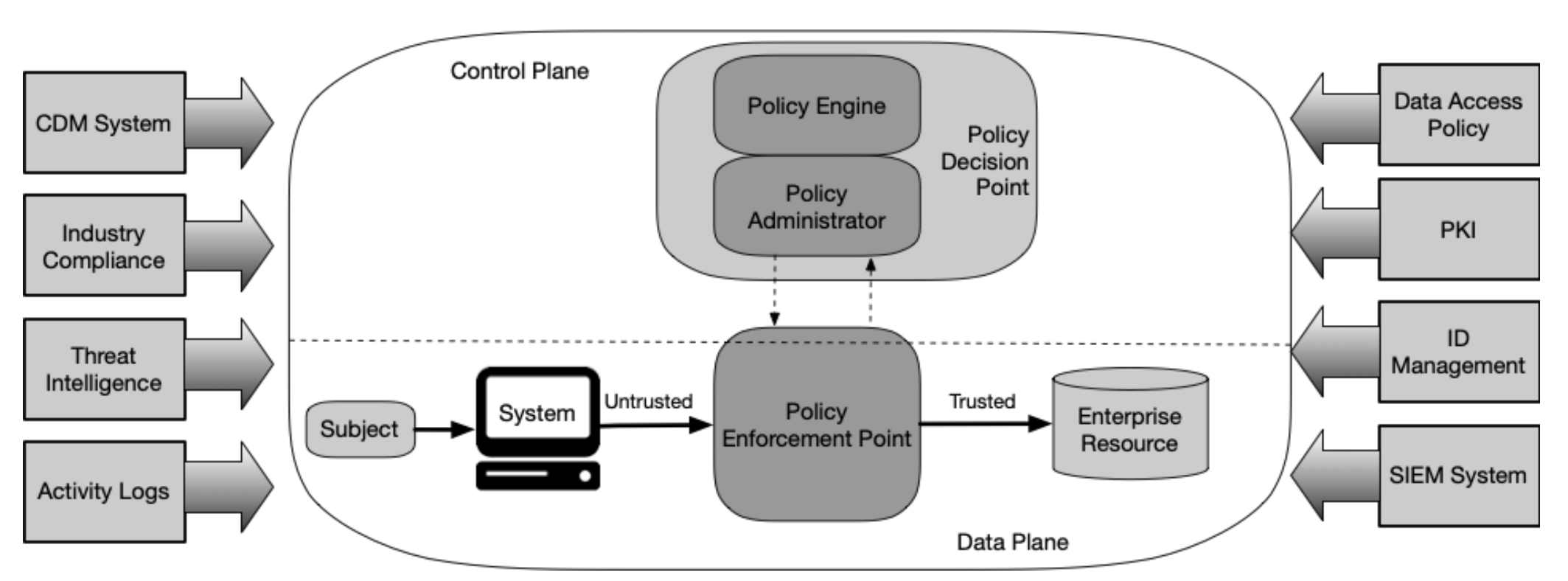}
\caption{Core Zero Trust Logical Components \cite{RBMC20}}
\label{fig:core_zerotrust_logical}
\end{figure}

To be more specific, ZTA requires the logical or physical separation of communication flows used for controlling and configuring the network from those used for actual organizational work as in Figure~\ref{fig:core_zerotrust_logical}. This means that the network must logically separate, i.e., introducing \emph{abstraction} between, the \emph{data plane} and the \emph{control plane}, as seen in the SDN design. The data plane handles communication between subjects and enterprise resources, and subjects should not be able to connect to enterprise resources without accessing the policy enforcement point (PEP). Access policies can be programmed at the policy decision point (PDP) in the control plane and enforced \emph{automatically} by communicating with PEPs in the data plane. This software-defined approach enables a central security control tower to gain visibility into access activities and \emph{automate} security policy configuration for tens of thousands of security devices.

\subsection{Software-Defined Approach and Crypto-Agility}

Since the path to zero trust is an incremental process that may take years to implement, 
CISA released the `Zero Trust Maturity Model' (ZTMM) \cite{CISA} to provide a guideline for the transition to zero trust. More specifically, the ZTMM suggests that gradual advancement, from the starting point \emph{traditional} to \emph{initial}, \emph{advanced} and \emph{optimal}, can be achieved over time across five distinct pillars: identity, devices, networks, application \& workloads, and data. Each pillar contains details about the cross-cutting capabilities, including visibility and analytics, automation and orchestration, and governance, and the maturity level of each pillar is determined by the level of optimizations of these capabilities. 

To achieve an optimal level of zero trust maturity, the zero trust architecture must be implemented diligently, guided by the design principles of software-defined approaches. It is important to note that the optimal level of the 'networks' pillar requires the integration of best practices for crypto-agility in the ZTMM. This suggests that the software-defined approach, characterized by abstraction and automation, contributes to the full implementation of both zero trust architecture and crypto-agility.

\section{Integrating PQC Migration into DevSecOps}

Considering the transition to post-quantum cryptography (PQC) as a software update, it is crucial to explore DevSecOps, which is the best practice for software updates in enterprise IT. This section highlights the importance of abstraction for effective implementation of cryptographic policy and demonstrates how DevSecOps can automate the cryptographic transition.

\subsection{Abstraction for Cryptographic Interfaces}

To ensure seamless software updates during the transition to new cryptographic standards, it is crucial to decouple static cryptographic configurations from applications. That is, applications should avoid directly invoking APIs of specific cryptographic modules or providers, as updates to cryptographic algorithms or standards may necessitate modifications. Java Cryptography Architecture (JCA) provides a standard interface, or \emph{abstraction layer}, that facilitates changes to cryptographic providers or suites without altering applications. This is achieved by updating the built-in \texttt{java.security} configuration file \cite{rwc2024}, typically found at \texttt{\seqsplit{\$jdk/jre/lib/security/}}. Additionally, specific algorithms can be enforced by defining policies within this file. Native applications, such as Nginx using OpenSSL, can similarly transition to a PQC-enabled TLS module by supporting a TLS plugin that is dynamically linked with OpenSSL.

\subsection{Automation of Cryptographic Policies Enforcement}

Despite the introduction of abstraction for cryptographic interface, manually updating the configuration file for cryptographic transitions remains necessary. This manual process is time-consuming and prone to errors, making it challenging to manage such configurations at scale in enterprise IT environments. The DevSecOps platform addresses similar issues by enabling the configuration of IT infrastructure and the enforcement of security policies at scale through code, such as Infrastructure as Code (IaC) and Policy as Code (PaC), within continuous integration and  continuous deployment and delivery (CI/CD) pipelines.

IaC automates the configuration and management of IT infrastructure, such as servers and networks, saving time and money while reducing human errors. Similarly, PaC allows security or compliance policies to be written as code for automatic execution. Both IaC and PaC ensure cost-efficient, consistent, and error-free IT operations with repeatability. A cryptographic administrator, or crypto officer, is then able to encode cryptographic policies, such as the migration to PQC algorithms, into PaC and save them in the PaC repository. IaC can then apply these updated policies automatically when launching applications or services, ensuring that the latest cryptographic configurations are consistently enforced.

\section{Towards Software-defined Cryptography}

\begin{figure}[ht] 
\centering
\includegraphics[width=\linewidth]{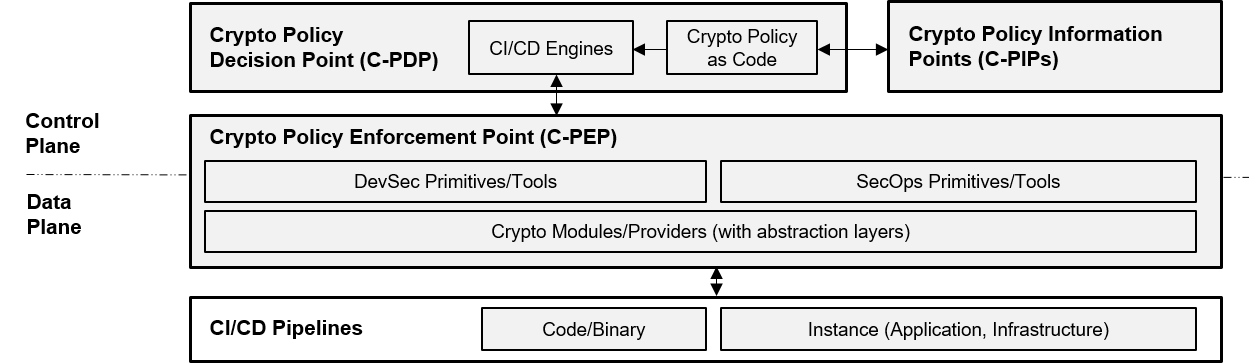}
\caption{Architecture of Software Defined Cryptography \cite{RBMC20}}
\label{fig:crypto_agility_architecture}
\end{figure}

We have investigated the design features of software-defined approach and DevSecOps methodologies, characterized by \emph{abstraction} and \emph{automation}, and now present the notion of software defined cryptography. Figure~\ref{fig:crypto_agility_architecture} depicts its conceptual framework which describes basic components and their interaction. The compliance or risk management framework from the Cryptographic Policy Information Point (C-PIP) is used to write \emph{cryptographic policies as code}, defining security policies related to the discovery or enforcement of specific cipher suites. The CI/CD engines in the Cryptographic Policy Decision Point (C-PDP) within the control plane invoke and \emph{automatically} deploy these policies to DevSec tools, SecOps tools, and cryptographic providers in the Cryptographic Policy Enforcement Point (C-PEP) in the data plane. Cryptographic providers and DevSec and SecOps tools should provide interfaces, such as \emph{abstraction} layers, to configure the use of specific cryptographic modules or algorithms. This design feature enables software-defined cryptographic policies to be enforced throughout CI/CD pipelines.

Enterprise applications are increasingly adopting a standardized architecture consisting of multiple \emph{loosely} coupled components called microservices, often deployed as containers. These applications are supported by an infrastructure that provides application services, such as a \emph{service mesh}. It embraces the software defined approaches, separating control plane for the purpose of management and configuration of application services, e.g., proxy within pods, in the data plane. This architecture enables application environment to be defined and managed with codes, including {application code} for business logic, \emph{application services code} for services such as session establishment, network connection, etc., \emph{infrastructure as code} to provision and configure compute, networking, and storage resources, \emph{policy as code} to define runtime policies such as zero trust, and  \emph{observability as code} to monitor application runtime state. This application environment, or architecture, enables efficient DevSecOps implementation, where development, deployment, and operation of the application can be agile and automated with elements such as continuous integration, delivery, and deployment (CI/CD) pipelines \cite{CHA22}. A service mesh with a fully-featured DevSecOps platform would fully support software-defined cryptography. In other words, cryptographic policies, including detecting weak cryptography or requiring specific cryptographic modules or algorithms, would be written as code and automatically implemented throughout the CI/CD pipelines.

\section{Conclusion}

The software-defined approach has become widespread in Enterprise IT, where functions such as networking and storage in a data plane are defined by software. This facilitates visibility and automation through a control plane that interacts with the data plane via well-defined interfaces. The paradigm of cybersecurity, exemplified by the zero trust architecture, is increasingly embracing the software-defined approach, and it is now time for cryptography to follow suit. This shift is particularly crucial as we face the challenge of migrating to post-quantum cryptography – a path that cryptography has not previously traversed.

\bibliographystyle{plain}
\bibliography{s-cape}

\end{document}